# "The constancy, or otherwise, of the speed of light"

Daniel J. Farrell & J. Dunning-Davies,
Department of Physics,
University of Hull,
Hull HU6 7RX,
England.

J.Dunning-Davies@hull.ac.uk

**Abstract.**

New varying speed of light (VSL) theories as alternatives to the inflationary model of the universe are discussed and evidence for a varying speed of light reviewed. Work linked with VSL but primarily concerned with deriving Planck's black body energy distribution for a gas-like aether using Maxwell statistics is considered also. Doubly Special Relativity, a modification of special relativity to account for observer dependent quantum effects at the Planck scale, is introduced and it is found that a varying speed of light above a threshold frequency is a necessity for this theory.

## 1. Introduction.

Since the Special Theory of Relativity was expounded and accepted, it has seemed almost tantamount to sacrilege to even suggest that the speed of light be anything other than a constant. This is somewhat surprising since even Einstein himself suggested in a paper of 1911 [1] that the speed of light might vary with the gravitational potential. Interestingly, this suggestion that gravity might affect the motion of light also surfaced in Michell's paper of 1784 [2], where he first derived the expression for the ratio of the mass to the radius of a star whose escape speed equalled the speed of light. However, in the face of sometimes fierce opposition, the suggestion has been made and, in recent years, appears to have become an accepted topic for discussion and research. Much of this stems from problems faced by the 'standard big bang' model for the beginning of the universe. Problems with this model have troubled cosmologists for many years; the horizon and flatness problems to name but two. The big bang was the fireball of creation at the moment this universe came into being. It had a temperature and, in keeping with thermodynamics, as it expanded its temperature dropped. If we model the big bang as a blackbody, we find that due to its temperature it emits most energy at a characteristic wavelength, figure 1a.

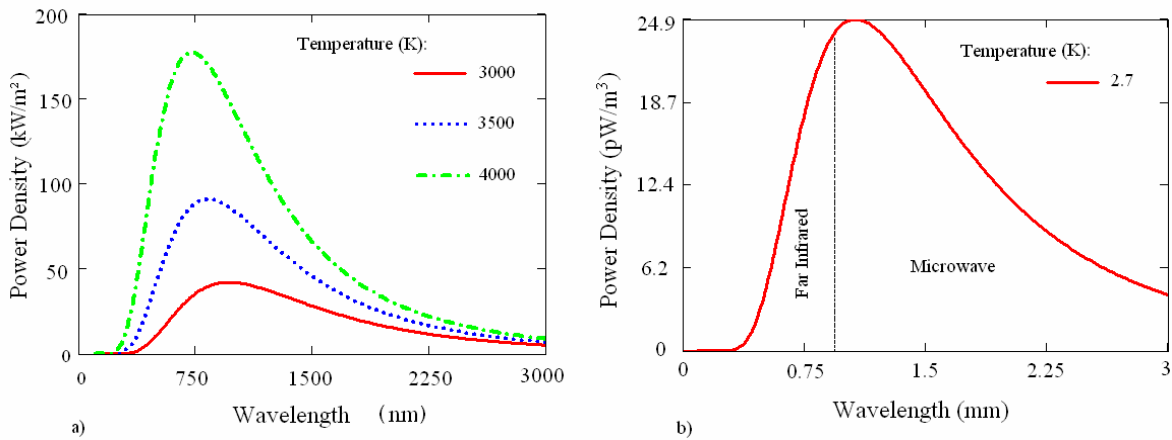

*Figure 1 – a) Theoretical blackbody radiation plots for 3 radiating bodies of different temperatures, with rate of energy in the kilowatt region. It can be seen that maximum energy is emitted at one wavelength, $\lambda_{max}$. b) Blackbody curve of the microwave background radiation. The observed isotropic radiation has a wavelength length of approximately 1mm (i.e. in the microwave region of the electromagnetic spectrum) and rate of energy in the picowatt region. The corresponding temperature of the 'body' is shown.*

The microwave background radiation ◊, or cosmic background radiation as it is sometimes called, is an isotropic radiation in the microwave region of the spectrum that permeates all of space. It is regarded as clear evidence for the big bang because, as shown in figure 1b, if $\lambda_{max}$ – which can always be measured - is known, the temperature of the emitting body can be calculated. When this idea is applied to the present day universe, it is found that it has a temperature of approximately 3K. This is the temperature of the universe, which is, in effect the present temperature of the cooled big bang explosion.

---

◊ Although initially discovered in 1940 by A. McKellar [3,4] it is generally erroneously accepted that Penzias and Wilson discovered the microwave background radiation in 1964.

If two regions of the sky that are both 13 billion light years away but in opposite directions are viewed, it is impossible for them to be in causal contact because the distance between them is greater than the distance light can travel in the accepted age of the universe, figure 2. This raises the question, 'why is the universe isotropic when all regions are in thermal equilibrium though they are not in causal contact'? This is the horizon problem.

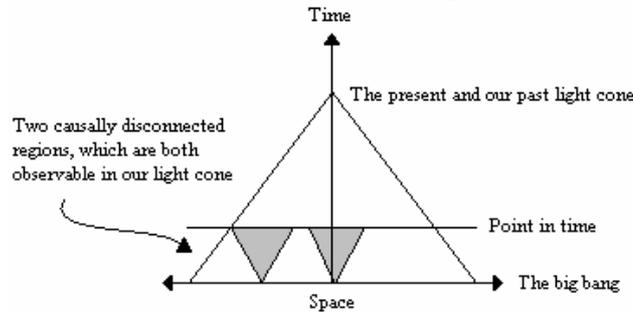

*Figure 2 – Horizon light cone due to the standard big bang model. Our past light cone contains regions outside each others' horizons.*[5]

Hubble discovered that the universe is expanding, namely that every point of space is moving away from every other point. But what is happening to the rate of this expansion? Cosmologists argue that this question depends on the matter content of the universe because the collective gravity could counteract the ballistic force of the big bang explosion. An analogy is that if a ball is thrown into the air, at some maximum height it will stop and then reaccelerate, in the opposite direction, back down to the thrower. The ball fell back for two reasons: the collective gravity of everything on earth was pulling it back and the ball was thrown with a velocity less than 11.2 km/s - the escape velocity.

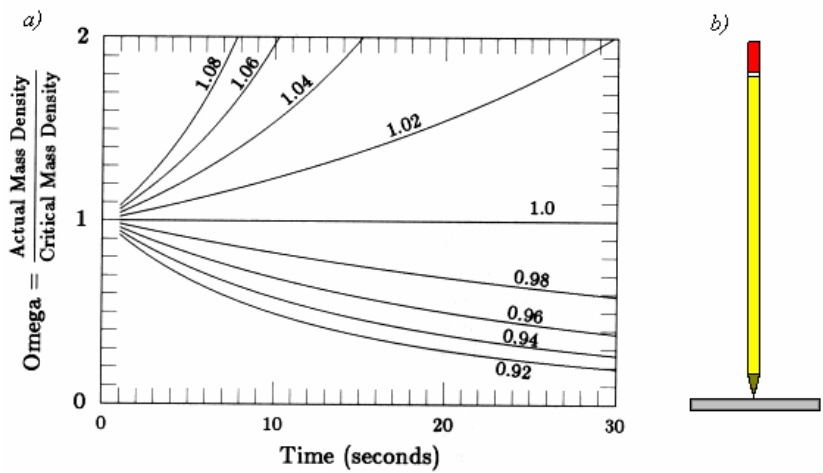

*Figure 3 – a) [6] Every line represents a potential universe each with a different starting value of omega. It can be seen how the omega of a given universe changes a function of time unless it has a value one. For example, the universe that has a starting omega of 0.98 diverges to omega of 0.6 just 30 seconds after the big bang. b) [6] This is an example of unstable geometry that underlies the flatness problem. The balance of the pencil is analogous to a universe of starting omega of 1 as any perturbation will disturb the system and rapidly cause it to move out of balance.*

Now imagine that it is possible to throw the ball considerably faster, up and beyond 11.2 km/s what will happen? If the ball is thrown with a velocity greater than 11.2 km/s, the ball will leave the gravitational pull of the earth and continue heading away forever, never coming to a standstill. An interesting case is when the ball is thrown *at* the escape velocity. Here the ball never truly escapes the gravitational pull; it will forever be moving away and will only come to a rest at infinity.

If this notion is developed to apply to the expansion of the universe it is seen that, if the gravitational attraction or 'mass-density' causing the gravitational attraction is at a critical level, the expansion will be halted giving a 'flat' universe. Cosmologists refer to the ratio of the actual density to this critical density by the Greek letter $\Omega$. From Figure 3a it can be seen that, if $\Omega$ is at any other value than one, the universe will rapidly take on an 'open' or 'closed' form; i.e. the ball leaving the earth's attraction or falling back down. Open and closed refer to a universe expanding forever or contracting to zero size, respectively, figure 4.

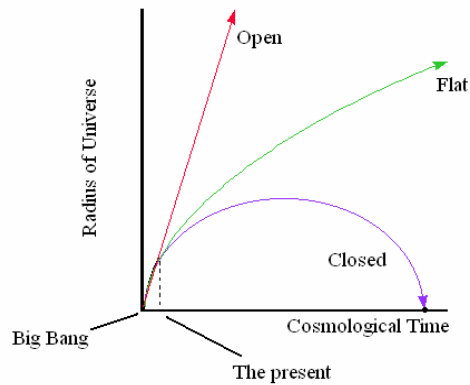

*Figure 4 – Fate of the universe will be either open, flat or closed depending on the mass-density omega.*

At present, $\Omega \approx 0.2$, which is incredibly close to the critical value, as from figure 3a it can be seen how fast omega will diverge away from one even ten seconds after the big bang, yet alone 15 billion years. For $\Omega \sim 0.2$ today, requires that $\Omega$ was 0.999999999999999 one second after the big bang, it seems difficult to explain how an explosion could be so fine tuned! An $\Omega = 1$ can be likened to the unstable geometry of a pencil balanced on its tip, figure 3b, as *any* perturbation would force it to rapidly fall into a more stable arrangement. This is the flatness problem. Inflationary theory, proposed by Guth in 1981 [7] has until now been the main contender in solving these problems. Inflation requires that, shortly after the big bang, expanding space experiences a period of superluminal expansion, that is, where space between two points expands faster than light can travel between those two points. This solves the horizon problem as our observable universe inflated from an elemental and, therefore, already homogeneous, volume of space. It also solves the flatness problem, imagine temporarily pausing the big bang expansion when it was a convenient size, say that of a football, an observer could clearly see that the space is curved. After inflation, the expansion is paused again, but now the observer believes that space is flat and Euclidean. However, objections to the original theory, though not necessarily to the basic idea of inflation, have been raised on thermodynamic grounds [8].

## 2. Varying speed of light theories (VSL).

In the last few years, alternative explanations for these problems have arisen. In 1993 John Moffat proposed a varying speed of light theory as a solution to the flatness and horizon problems [9]. Although Moffat is a prolific theoretician, his initial paper received little recognition until in recent years when a team from Imperial College London - Andrew Albrecht and João Magueijo - reintroduced the ideas [5]. There are many forms of VSL theory as, at first sight, it is a relatively new area of research with many different teams working on it from around the world.

Magueijo investigated what effect a variable speed of light would have on different areas of physics, his initial approach was by adding '$c$-dot-over-$c$'[℘] correction terms to various standard physics formulae. However, as modern physics has been built up using a constant speed of light this soon became a daunting prospect, but at the same time an 'embarrassment of riches'.

Perhaps the most controversial result from VSL is the violation of energy conservation. This is realised by both Moffat and Magueijo who deal with it in separate ways. In fact, it is hard to see how a varying speed of light could not violate energy conservation, as energy is related to mass via $E=mc^2$[#]. Magueijo meets this problem head on with the opinion that, by assuming energy conservation you have already assumed the speed of light to be constant, he sums it up as:

> '...the conservation of energy is simply another way of saying that the laws of physics must be the same at all times...' [10]p156

Presumably, this is because, if the laws of physics change, then it is highly probable that the energy associated with a system or interaction will also change. Therefore, VSL intrinsically disobeys energy conservation.

To reject VSL theory outright just because it disagrees with the conservation of energy would be closed minded. Also, if nobody was allowed to publish articles that disagree with present theories, there would be little forward development in physics. However, energy conservation is something which is observed everyday in our lives. How can it be challenged?

### Magueijo's VSL.

In general relativity, the presence of matter and energy is thought of as 'curving' space-time. This central tenet can be stated in the innocuous looking equation:

$$G^{ij}=kT^{ij}, \qquad (1)$$

where $G^{ij}$ is the 'Einstein tensor', which encapsulates all the information regarding the geometry of space-time. Einstein made the assumption that this geometry is proportional to $T^{ij}$, the 'energy-momentum tensor', containing information about the matter and energy in that space. However, he needed a proportionality constant, $k$, to connect the two.

It is well known that Newton's theory of gravitation agrees exceptionally well with experiment. Therefore, a proviso of General Relativity must be that, under appropriate assumptions, it will lead to a so-called 'Newtonian approximation'. In such a situation, weak

---

[℘] '$c$-dot-over-$c$' refers to a mathematical shorthand of explaining rates of change with time. i.e. the ratio of the rate of change of the speed of light with time and the speed of light.

[#] It could be argued that this equation was derived by Einstein using a constant speed of light and so the above comparison may not be drawn. However it is possible to derive the energy-mass relationship without using relativistic physics as demonstrated by Poincare (1900)[11] and Born [12].

gravitational fields and low velocities are assumed. When this is done, it can be shown such a proportionality constant comes out to be [13]:

$$k = 8\pi G/c^4 \quad (2)$$

It is important to note that the speed of light always appears in the value of '$k$' – even when assuming little curvature, i.e. the Newtonian approximation. Magueijo's reasoning is that '$k$' involves '$c$' (which is no longer a constant) and so, '$k$' is not a constant and, since '$k$' relates to what degree space-time is warped due to the presence of matter and energy, there must be an interplay between the degree of curvature and the speed of light in that region.

It is known that a universe with $\Omega > 1$ is closed, where the energy-density must be large. Taking the energy-momentum tensor, $T^{ij}$, from equation (1) and for simplicity's sake just looking at its energy component it may be written:

$$T^{ij} = mc^2 \quad (3)$$

Combining equations 1, 2 and 3 gives:

$$G^{ij} = (8\pi G) \cdot m/c^2 \quad (4)$$

It follows from equation (4) that, if '$c$' increases, the right-hand-side will decrease. This must have the effect of reducing the related term in the Einstein tensor, $G^{ij}$, hence reducing the curvature of space-time and pushing the universe away from its closed fate. Not only does this occur, but by actually reducing the curvature, the energy-density is being actively reduced. This is indicative of the homeostatic properties of coupled differential equations, which is what equation (4) represents.

The inverse is also valid; namely, that an open universe with $\Omega < 1$ will have an energy-density lower than unity. Hence, the speed of light will decrease, resulting in an increase in energy-density and a pushing of $\Omega$ towards one. VSL then implies a flat universe, as any change in energy-density away from unity is seen to result in an action to pull it back to unity or as Magueijo writes:

> '*Under our scenario, then, a flat universe, far from being improbable, [is] inevitable. If the cosmic density differed from the critical density characterising a flat universe, then violations of energy conservation would do whatever was necessary to push it back towards the critical value.*' [10]p158

It can be seen how this explains why the universe is so nearly flat and also, how a theory, whose initial premise is to violate energy conservation, gives a result that is in keeping with energy conservation. It attempts to make the 'energy-gradient' of the universe zero and, therefore, negating the need for the speed of light to change. One could also argue that this energy gradient, associated with the curvature of space, is so incredibly small over the scale at which experiments can be conducted that it has no effect. Indeed, on this scale, Einstein's principle of equivalence could be approximated to:

> "*Gravitational fields may always be transformed away in an infinitely small region of space-time.*"

VSL also solves the horizon problem by assuming – much like inflation – that, at some instant after the big bang, a change took place to the universe; in VSL it is argued that this is a 'phase transition'. Before this transition light could travel approximately thirty orders of magnitude faster, thereby allowing all the universe to be in causal contact, figure 5.

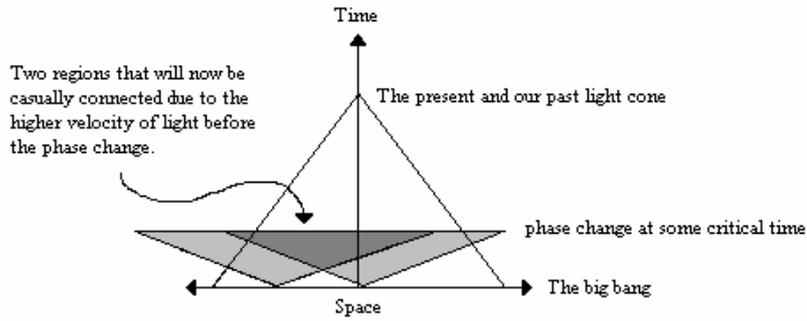

*Figure 5 – The horizon of the light cones in the early universe is much larger than before; this is due to the higher velocity of light before the phase change.*

**Moffat's VSL**

In his 1993 paper [9], Moffat applies VSL specifically to the initial conditions of the big bang and proposes it with a view to directly replace inflation. He argues that one problem with the inflationary model is that it contains certain 'fudge-factors'; namely the fine tuning of potentials that operate the inflationary expansion. Allowing that this is virtually a prerequisite for any theoretical exercise, it is of concern because the more factors that have to be introduced 'manually' imply a poor underlying theory. Moffat argues that, although the process of inflation provides a method for solving cosmological problems, to bring it in line with observation requires some tweaking. This involves multiplying various assumed potentials, which operate on the expansion, by tiny numbers ($10^{-12}$). This is done in order to correct the 'nucleation rate of bubbles' or inhomogeneities in the early universe that were the seeds for galaxies.

Moffat also finds contradictions in various inflation models. For example, in 'Linde's Chaotic Inflation' it is necessary to fine tune certain 'coupling constants' to be very small numbers (in the region of $10^{-14}$). This is needed to bring the model into line with the present observed density profile. However, there is no physical reasoning behind this; it is just a mathematical pursuit. Moffat also notes that by forcing this result, it has the consequence of producing results 'which are not in keeping with the original ideas of inflation…as the theory's potential is now uniform over a region greater than the Hubble radius' [9]. Clearly, some inflationary models do not appear to recognise that the mathematics must have a direct connection with physical reality.

Moffat states that VSL requires less of this tweaking:

> *"…superluminal model could be an attractive alternative to inflation as a solution to cosmological problems."*

In his papers [9,14] Moffat gives a very detailed mathematical description of why this is so by showing how a varying '$c$' could modify the big bang to give today's universe. Moffat states that the empirical basis for varying '$c$' comes from 'broken symmetry phase changes' in the early universe.

This is the concept that the separate forces we observe as gravity, electromagnetism and the nuclear forces are actually manifestations of one underlying force. In the early universe, when the average thermal energy of a particle was in the region of $10^{19}$ GeV, ~particles are thought to have experienced this single force.

---

~ Particle physicists measure the mass of particles in the convenient units of energy because at the nuclear level all forms of energy can be transmuted into mass and visa-versa. Moreover, by using units of eV (electron-volts) we can treat a particle possessing a rest mass, relativistic mass and kinetic energy by one number.

A theory based on the notion of unifying forces was proposed by Glashow, Weinberg and Salam in the 60's. They demonstrated theoretically that, at high energies, the electromagnetic force – itself a unification of the electric and magnetic forces – is unified with the weak nuclear force forming the 'electro-weak force'. Experiments at CERN in 1972 showed evidence supporting this theory by finding the postulated neutral force 'exchange particle'.

From the standard model of particle physics, forces are represented by 'exchange particles'. For example, the exchange particle of the electromagnetic force is the photon (light). Photons carry or exchange electric and magnetic fields of force through space. However, the photon is a massless particle, but the exchange particles for the weak interaction – the W and Z Bosons – have masses of approximately 80 GeV$^\zeta$. This implies that, at least equilibrium energy, is needed for the particle to be freely observed. When the equilibrium energy dropped to below this energy, the electro-weak force separated into the electromagnetic and weak nuclear forces - the symmetry of the initial unification lost.

It can be seen, using an analogy, what such a dominating effect this spontaneous breaking of symmetry must have had. It is known that the different phases of water: solid, liquid and vapour occur when there are varying levels of energy available to a group of water molecules. When little energy is available the intermolecular forces pull the molecules together. The hydrogen in the molecule orientates itself to be as close as possible to the oxygen in other water molecules forming ice. At a critical energy level or temperature, the molecules suddenly acquire sufficient energy to overcome this bond, forming a loosely bound liquid state. At yet another critical point, the molecules gain sufficient energy to disassociate entirely, forming steam. If heating continues then, at another critical energy level, the electrons orbiting the molecule will have sufficient energy to leave their bound state, breaking the bond between the hydrogen and oxygen forming a plasma or ionised gas.

The phases of matter are then representative of some underlying order in the system and where there is order there is also symmetry. Symmetry breaking then implies a massive upheaval in the internal ordering of a system.

In the case of the early universe, as the average energy level passed through various critical values, which correspond to the energy associated with respective exchange particles, forces decoupled to appear as if they were separate, figure 6.

---

$^\zeta$ The thermal energy or equilibrium energy of a group of particles is distributed around a mean value given by $kT$, where $k$ is Boltzmann's constant and $T$ the absolute temperature. The thermal energy of a particle at room temperature is 0.025 eV some 12 orders of magnitude smaller in difference.

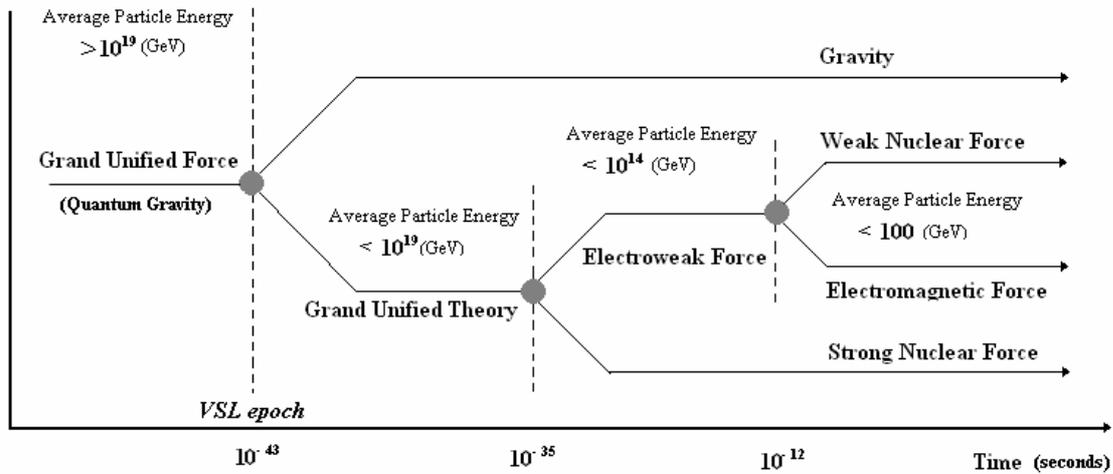

*Figure 6 – Approximate time and energy for each epoch as the unified force de-couples and the proposed varying speed of light epoch.* [15]

At each point the universe arranged itself in the best symmetry, or order, possible at the time. These phases are often referred to as 'epochs' by cosmologists to reinforce the huge changes taking place. In the 'VSL epoch', so called 'Lorentz invariance is spontaneously broken'; meaning that the Lorentz transformations, or more specifically Einstein's second postulate: 'The speed of light $c$ is the same constant with respect to all observers irrespective of their motion and the motion of the source' no longer applies. Moffat then assumes that at some critical temperature, $t_c$:

> '...[a] phase transition generates a large increase in the speed of light and a superluminary communication of information occurs, allowing all regions in the universe to be causally connected...the conservation of energy-momentum is spontaneously violated and matter can be created in this broken symmetry phase..' [9]

Moffat suggests a time frame for the VSL symmetry breaking to occur:

> 'The symmetry breaking will extend to the singularity or the possible singularity-free initial state of the big bang, and since quantum effects associated with gravity do not become important before $t \sim 10^{19}$ GeV, we expect that $t_c \geq 10^{19}$ GeV.' [9]

This is because, from figure 4, it is seen that, when the average particle energy is $10^{19}$ GeV, or alternatively, $t_c \geq 10^{19}$, quantum gravity is no longer required to describe the space, it can be described by general relativity.

In his theory, Moffat describes the early universe by starting with an equation that defines that space, the line element.

$$ds^2 = dx^2 + dy^2 + dz^2 \qquad (5)$$

Equation (5) is the standard Euclidean/Newtonian line element. It gives the position of any point in space relative to the origin of the $x, y, z$ axes. In special relativity, space and time are no longer separate they are linked by the definition of the line element used:

$$ds^2 = c^2 dt^2 - dx^2 - dy^2 - dz^2 \qquad (6)$$

In situations that have spherical symmetry it is common to use a different coordinate system. Rather than using three distances along orthogonal axes, a point in space can be defined by two angles and a length. If equation (5) is rewritten in spherical polar coordinates,

$$ds^2 = r^2 + r^2 d\theta^2 + r^2 \sin^2\theta d\varphi^2 \qquad (7)$$

results. Moffat uses a result by Robertson and Walker which is a modified spherical polar line element to include $R(t)$ and $k$:

$$d\sigma^2 = R^2(t)\left[\frac{dr^2}{1-kr^2} + r^2(d\theta^2 + \sin^2\theta d\phi^2)\right] \quad (8)$$

where $k$ is a re-worked omega so that an open, flat and closed universe makes $k$ equal to 1, 0 or -1 respectively and $R(t)$ is the radius of curvature of the universe as a function of time. Treating equation (8) equal to equation (7), it may be inserted into the spherical polar version of equation (6) to give the Friedman, Robertson, Walker line element:

$$ds^2 = dt^2c^2 - R^2(t)\left[\frac{dr^2}{1-kr^2} + r^2(d\theta^2 + \sin^2\theta d\phi^2)\right] \quad (9)$$

Equation (9) describes a space consistent with special and general relativity that is isotropic and homogeneous. This equation may be modified to include a varying speed of light. This is allowed under the assumption that during the spontaneous phase transition Lorentz invariance is broken. Moffat uses 'Heaviside step functions' to allow the speed of light to increase and decrease discontinuously. Equation (9) then becomes,

$$ds^2 = dt^2c^2(t) - R^2(t)\left[\frac{dr^2}{1-kr^2} + r^2(d\theta^2 + \sin^2\theta d\phi^2)\right] \quad (10)$$

where $c(t)$ is given by,

$$c(t) = c_0\theta(t_c - t) + c_m\theta(t - t_c) \quad (11)$$

where $\theta$ is the Heaviside step function, $c_o$ the speed of light before the transition and $c_m$ (which is known to be $3 \times 10^8$ ms$^{-1}$) the speed of light after the transition. This gives $c(t)$ the property to decrease suddenly in value at the critical time, $t_c$.

Moffat's theory is biometric; that is, space is defined by two different metrics or line elements, one for $c_o$ and one for $c_m$ [14]. Two light cones are necessary for every point in space, where the relative size of the cones is given by:

$$\gamma = c_o/c_m \quad (12)$$

The theory explains the horizon problem because the $c_o$ light cone can become much larger than the $c_m$ light cone, see figure 4. This is the same concept as discussed before, but with the mathematical backing.

Moffat demonstrates how VSL solves the flatness problem by writing the Friedman equation at the point of broken symmetry. It is shown that, to acquire a mass-density in keeping with observation, $\Omega \sim 1$, and also, for a speed of light after the transition to be $3 \times 10^8$ ms$^{-1}$, it is a necessary condition that the early speed of light be $\sim 1.5 \times 10^{37}$ ms$^{-1}$; an increase, from today's velocity of twenty-nine orders of magnitude. Hence, Moffat has shown that VSL solves the horizon and flatness problem and requires much less tweaking of critical parameters than current inflationary theories.

**Comparisons and Differences.**
These two theories differ in that Magueijo's VSL proposes a continuous mechanism that will eventually push the universe to $\Omega=1$, i.e. violations of energy conservation; Moffat's theory, however, is aimed as an alternative to the inflationary model of the universe, setting the initial conditions so that the observed universe emerges.

Although both theories require the speed of light be greater in the early universe, they have different reasons. The curvature of space-time in the early universe would have been very great. Hence, from Magueijo, the speed of light is required to be greater than ~$3 \times 10^{38}$ ms$^{-1}$ [5] and from Moffat, symmetry breaking in the early universe provides a mechanism for a change in '$c$' through breaking Lorentz invariance, thereby allowing the speed of light to become ~$1.5 \times 10^{37}$ ms$^{-1}$ [14]. It is interesting that these two mechanisms for varying '$c$' predict a change in '$c$' of approximately the same amount in the early universe. It could be that these two versions of VSL theory are elements in a more deep rooted theory of the universe.

### 3. Thornhill and VSL.

It is possibly of interest to note that the work of Moffat came to light because of a note added in proof to the article by Albrecht and Magueijo. However, the work of Thornhill remains largely unknown. This is due in part to his being forced to publish in somewhat obscure journals because he remains skeptical about the validity of the theory of relativity. However, in 1985, he published an article in *Speculations in Science and Technology*, entitled "*The kinetic theory of electromagnetic radiation*" [16]. In this article, it was shown that Planck's energy distribution for black body radiation may be derived simply for a gas-like aether using Maxwell statistics. Over the years, countless reasons have been put forward denying the existence of an aether. One asserts that Maxwell's equations show electromagnetic waves to be transverse and so, any aethereal medium must behave like an elastic solid. However, Maxwell's equations show only that the oscillating electric and magnetic fields are transverse to the direction of wave propagation; they say nothing about any condensational oscillations of the medium in which the waves are propagating. Hence, the assertion is incorrect. As Thornhill pointed out, if such a medium does exist, since the electric field, the magnetic field and the motion are mutually perpendicular for plane waves, Maxwell's equations would lead to the conclusion that the condensational oscillations of the medium are longitudinal, just as occurs with sound waves in a fluid. A second argument claims that Planck's formula for the black body energy distribution cannot be derived from the kinetic theory of a gas using Maxwell statistics. Actually, kinetic theory and Maxwell statistics lead to an energy distribution which is a sum of Wien-type distributions for a mixture of gases with any number of different kinds of particle. It is pointed out that this merely states the impossibility of so deriving the Planck distribution for a gas with a finite variety of particles. Hence, to complete the alleged proof, it would be necessary to eliminate the case of an infinite variety of particles. Obviously, this would refer to 'infinite' in a mathematical sense and this, in physical practice, would mean a very large number, as is so often the case in present-day physical theories. Thornhill goes on to derive the Planck distribution for a gas consisting of an infinite variety of particles whose masses are integral multiples of the mass of the unit particle. In this case, the frequency of the electromagnetic waves is found to correlate with the energy per unit mass of the particles, not with their energy; indicating a departure from Planck's quantum hypothesis. The special wave-speed, usually termed the speed of light, is identified with the speed in the background radiation and leads to a mass of $0.5 \times 10^{-39}$ kg for the unit aether particle. More importantly for the present discussion, he deduces that the speed of light must vary; in fact, the speed of light is found to vary as the square-root of the background temperature.

Obviously, as the title of his article indicates, Thornhill was more concerned with deriving the Planck formula and finding this dependence of the speed of light on the value of the background temperature came as an added bonus. This could be another reason his work was not recognized. However, it might be pointed out that his work was discussed at a conference held at Imperial College, London in 1996 [17] and, at that conference, it was explicitly stated that, if the speed of light did vary in the way suggested, it would offer an alternative explanation for problems dealt with at that time by the inflationary theories.

## 4. The Evidence for VSL.

It has been seen that VSL appears to solve a wealth of cosmological problems and does so in a simpler manner than conventional theories. However, other than its theoretical successes, is there any physical evidence for a varying speed of light? Webb et al [18] and Murphy [19] et al studied the spectra of quasars, extremely distant objects which emit immense amounts of power and are also strong X-ray sources. A typical quasar spectrum is shown in figure 7. Due to their great distance, heavily red-shifted spectra are observed.

Line spectra represent energy levels in atoms. As hydrogen on the other side of the universe is the same hydrogen that can be studied in the laboratory, it is known what wavelengths it adsorbs. Therefore, there is a reference to tell by how much a spectrum has been redshifted.

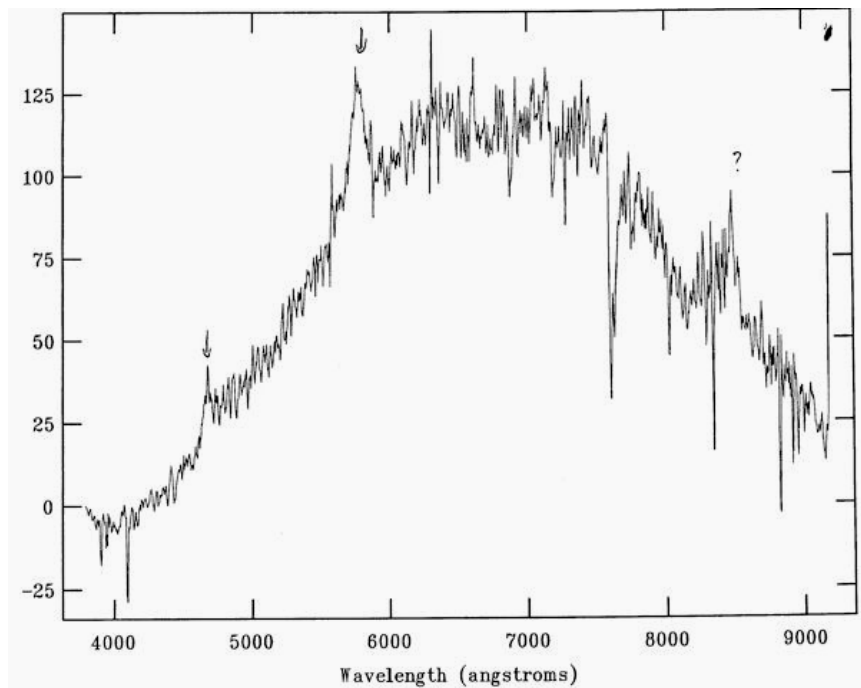

*Figure 7 – 'Standard' quasar spectra; intensity vs wavelength. Note the fine asorption lines and broad emssion lines indicate a quasar.* [20]

Fine-structure is the splitting of spectral lines into doublets, triplets etc. However, this is only noticeable at very high resolution because the separation of the lines is very small. It can be shown with the Bohr atomic model that the energy separation of the fine structure lines is:

$$\Delta E := \alpha^4 m \cdot c^2 \left(\frac{1}{n^5}\right) \tag{13}$$

where '$m$' is the mass of an electron, '$c$' is the speed of light, '$n$' is the primary quantum number and $\alpha$ is the fine structure constant.

The fine structure constant is one of four 'coupling constants' that describe the relative strength of the forces, see table 1.

| Force | Symbol | Dimensionless Value |
|---|---|---|
| Strong Nuclear | $\alpha_s$ | 1 |
| Electromagnetic | $\alpha$ | 1/137 |
| Weak Nuclear | $\alpha_w$ | $10^{-6}$ |
| Gravity | $\alpha_g$ | $10^{-39}$ |

*Table 1 – Values for fundamental coupling constant for the forces of nature.* [21]

The fine structure constant relates how electromagnetic radiation interacts with matter and has the form:

$$\alpha = (e^2/4\pi\varepsilon_0)(1/hc) \tag{14}$$

The first bracket is the coulomb force between two electrons; the second bracket is equal to the reciprocal of the product of the energy and wavelength of a photon [ς]. It may be shown that both brackets have the same units and so, '$\alpha$' is dimensionless; i.e. it is just a number.

As expected, quasar spectra show a redshift. However, there is an additional observation; the relative positions of the absorption lines change, i.e. they fan out, see figure 8.

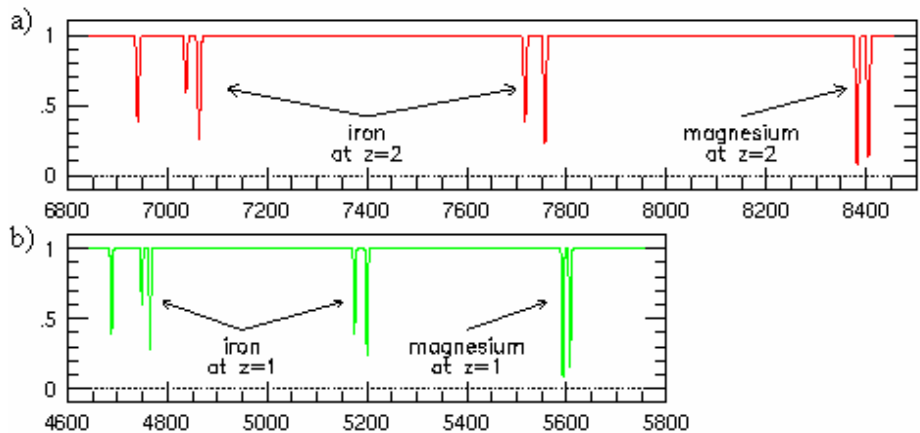

*Figure 8 – Theoretical plots of absorption spectra with a) redshift of 2 and b) redshift of 1. Note the redshift also has the effect of making the spectra fan out. This is taken as possible evidence for a time varying fine structure constant and consequently a time varying speed of light.* [22]

From Hubble's Law it is known that the velocity of recession - and, therefore, the redshift - increase with distance. It follows that the larger the redshift obtained for an object, the further back in space and time it is. Therefore, figure 9 shows a time varying fine structure constant that

---

[ς] The product of energy and wavelength this is derived from the de Broglie hypothesis: $E = hc/\lambda$ therefore $E\lambda = hc$.

was smaller in the early universe and is consistent with VSL theories due to '$c$' being inversely proportional to '$\alpha$'.

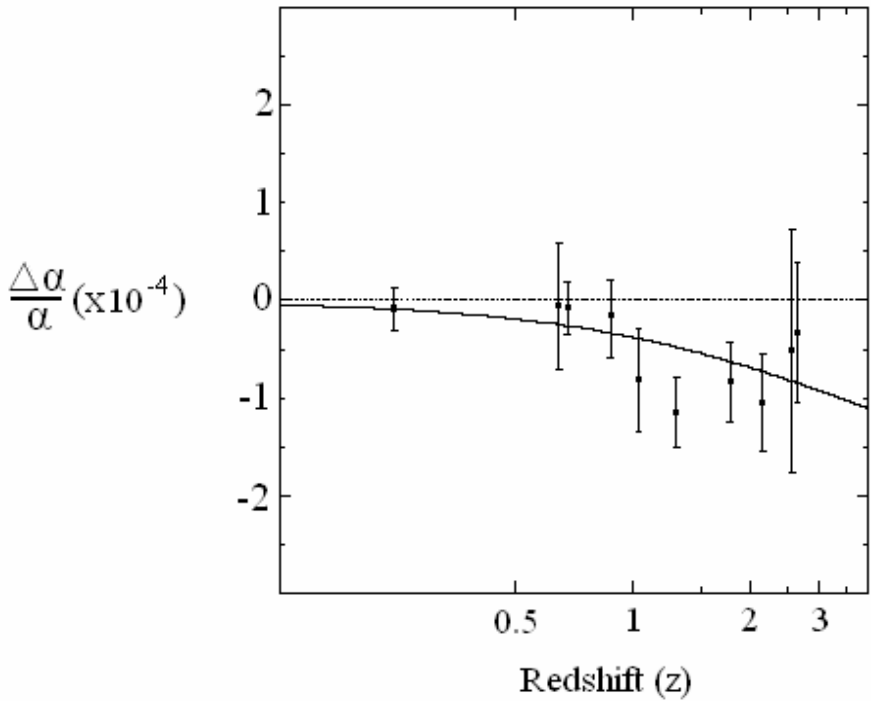

*Figure 9 – Quasar spectra observations, arranged as to corresponding to their apparent fine structure value plotted as a function of their redshift. Variation in the fine structure constant with time is implied. A theoretical 'varying-α' produced by a VSL theory is also shown. [23]*

Ultra-high energy cosmic rays are presumed to be protons or alpha particles of very high energies $10^{18}$ eV. When they hit the atmosphere, a shower of particles is caused. By finding the angle of the incident shower, counting the number of particles and measuring their energy, it is possible to work backwards and find the energy of the incident cosmic ray. They are also exceedingly rare observations, see table 2.

| Arrival Rate | Energy (eV) | Source |
|---|---|---|
| 1000 per meter$^2$ per second | $10^{10}$ | Milky Way Galaxy |
| 1 per meter$^2$ per second | $10^{12}$ | Milky Way Galaxy |
| 1000 per meter$^2$ per year | $10^{15}$ | Milky Way Galaxy |
| 1 per kilometer$^2$ per year | $10^{19}$ | Extra-galactic? |

*Table 2 - Approximate frequency of occurrence and energies of comic rays with their assumed source. [22]*

However, there is no local (in the galactic sense) source of protons with energy $10^{18}$ eV; they are then assumed to be 'extra-galactic'. Hence, they have been travelling through space with ultra-high energy for billions of years. This poses a problem. As discussed before, the microwave background radiation is isotropic and due to the sparseness of matter in intergalactic space, this is all the ultra-high energy cosmic rays will see. It isn't immediately apparent why this is a problem. However, a $10^{18}$ eV proton is a relativistic particle travelling with a velocity just short of the speed of light. From its rest frame, the cosmic background radiation photons, of 0.001 eV, look like gamma rays of energy $10^9$ eV [24].

It is known from experiments that matter/photons have a high cross section (interaction rate) with gamma rays. Therefore, interactions with the cosmic background radiation will absorb energy and slow ultra-high energy cosmic rays. It is as if space appears opaque to them, meaning that they have an average path length short enough to ensure that energies of ~$10^{20}$ eV and above can never be observed on earth. This is the Greisen-Zatsepin-Kuzmin limit, figure 10, and can by derived by considering relativistic arguments.

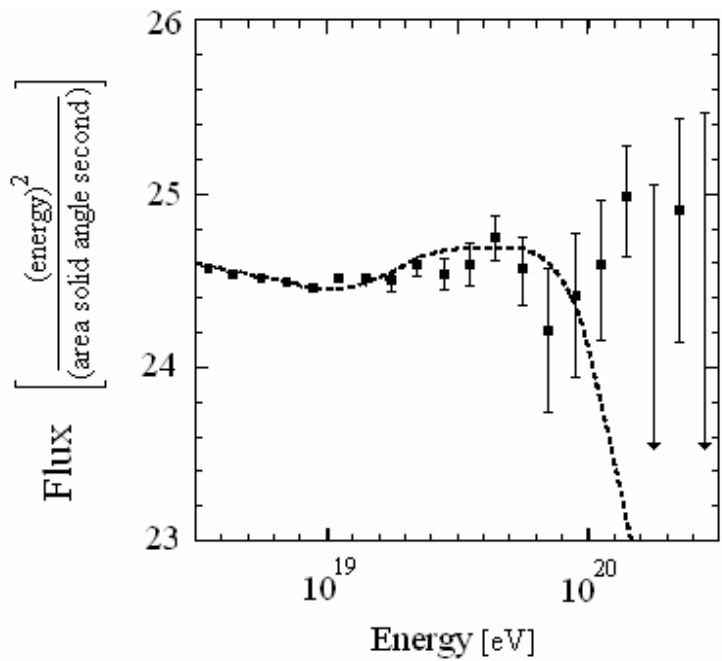

*Figure 10 – The Greisen-Zatsepin-Kuzmin limit for ultra-high energy cosmic ray observation and data points for observed ultra-high energy cosmic ray events; some rays are clearly above this limit. Flux is a measure of the flow of energy through an area. Cosmic ray data is uses area solid angle due to the geometry of the problem i.e. the cosmic ray cause a shower of particles in a cone shape to fall onto earth. [25]*

4.
5.   **5. Doubly Special Relativity.**

Giovanni Amelino-Camelia, of Universita' La Sapienza in Rome[24], has shown that the anomalies in cosmic ray energy may be explained by adding an extra postulate to relativity. According to quantum gravity theorists, the Planck length is the scale at which the 'granularity' of space-time due to quantum effects becomes dominant at around $4 \times 10^{-35}$ m (twenty orders of magnitude smaller than the radius of a proton). Planck scales exist also for other fundamental properties; mass ($m_p$) and time ($t_p$). Therefore, additional Planck scales may be defined; for example, the Planck momentum, $p_p$, is the product of $m_p$ and the 'Planck velocity' i.e. $\lambda_p/t_p$. It is also possible to define a Planck energy, $E_p$ ~ $10^{28}$ eV. Amelino-Camelia proposes:

*'Various arguments lead to the expectation that for particles with energy close to $E_p$ it would be necessary to describe space-time in terms of one form or another of [a] new space-*

*time…[for] our readily available particles, with energy much smaller that $E_p$ the familiar classical space-time picture would remain valid.'* [24]

Moreover, the energy is related to the de Broglie wavelength of a particle. If the energy is big enough so that the length approaches the Planck length, how will two different observers view the particle?

Due to the relative motion of observer and particle, the particle's length will contract. It is possible for a situation to arise wherein one frame of reference the particle is viewed as above the Planck length and from another frame of reference is viewed as below the Planck length. The two different observers would require different physics to describe the particle, therefore breaking Einstein's first postulate that the laws of physics are the same for every observer.

The concept that relativity is based on a velocity observer-independent scale, i.e. light is the only object in the universe to be allowed to have a constant speed from all frames of reference, is familiar to all. Amelino-Camelia proposes the introduction of an additional observer-independent scale based on the Planck length, $\lambda_p$ and momentum $p_p$. He restates the relativity postulates as [24]:

1) The laws of physics involve a fundamental velocity scale based on '$c$' and a fundamental length scale based on '$\lambda_p$'.
2) The value of the fundamental velocity scale '$c$' can be measured by each inertial observer as $\lambda/\lambda_p \rightarrow \infty$ limit of the speed of light of wavelength $\lambda$.

Moreover, photons of low energy travel at '$c$' while photons above a threshhold energy can have varying values, faster than '$c$', which are proportional to their energy. When the Greisen-Zatsepin-Kuzmin limit is re-derived to include this dependence of threshold energy, it is found to account for the anomalous cosmic ray results [26].

Doubly Special Relativity also provides a 'natural' mechanism for the speed of light to be faster in the very early universe, as the average energy per particle would have been well over the threshold value, figure 6.

## 6. Conclusions

Cosmologists are currently hunting for 'dark-energy', a mysterious substance that supposedly drives the current acceleration in the expansion of the universe. However, this is already predicted by some VSL theories, as discussed in 2.2.

For accepted cosmological theories to be valid, it is required that the universe is composed of 5% ordinary matter, 25% dark matter and 70% dark energy. It seems more realistic to believe in a varying speed of light via the mechanisms discussed above, rather than invent abstract conceptions simply because they happen to balance familiar cosmological equations.

If one day, realistic alternatives such as VSL are shown to be wrong, then alternative ideas should be suggested and experimentally proved. However, it seems that VSL is currently being over-looked by many scientists for its supposedly heretical ideas. However, in the words of Paul Dirac:

*"It is usually assumed that the laws of nature have always been the same as they are now. There is no justification for this. The laws may be varying with cosmological time."* [27]

Indeed, here the idea of, and evidence for, the change in one variable, '$c$', the speed of light has been discussed. However, theories have suggested that the electronic charge '$e$' [28] can also vary and work has also been done to suggest that the gravitational constant '$G$' is varying with cosmological time [29, 30].

Ultimately, theories will come and go but there will always be a slow movement in the direction of progress and truth. Or in the words of Robert Frost:

*'We dance in a circle and suppose, while the secret sits in the centre and knows.'*
**Robert Frost, The Secret Sits**